\documentclass[aps,prl,twocolumn,superscriptaddress,showpacs,amsmath,amssymb]{revtex4}

\usepackage{graphicx}
\usepackage{bm}
\usepackage{amssymb}
\usepackage{hyperref}

\begin{document}

\title{Observation of the Larmor and Gouy Rotations with Electron Vortex Beams}


\author{Giulio Guzzinati}
\affiliation{EMAT, University of Antwerp, Groenenborgerlaan 171, 2020 Antwerp, Belgium}

\author{Peter Schattschneider}
\affiliation{Institut f\"ur Festk\"orperphysik, Technische Universit\"at Wien, A-1040 Wein, Austria}
\affiliation{University Service Centre for Electron Microscopy, Technische Universit\"at Wien, A-1040 WIEN, Austria}

\author{Konstantin Y. Bliokh}
\affiliation{Advanced Science Institute, RIKEN, Wako-shi, Saitama 351-0198, Japan}
\affiliation{A. Usikov Institute of Radiophysics and Electronics, NASU, Kharkov 61085, Ukraine}

\author{Franco Nori}
\affiliation{Advanced Science Institute, RIKEN, Wako-shi, Saitama 351-0198, Japan}
\affiliation{Physics Department, University of Michigan, Ann Arbor, Michigan 48109-1040, USA}

\author{Jo Verbeeck}
\affiliation{EMAT, University of Antwerp, Groenenborgerlaan 171, 2020 Antwerp, Belgium}

\begin{abstract}
Electron vortex beams carrying intrinsic orbital angular momentum (OAM) are produced in electron microscopes where they are controlled and focused using magnetic lenses. We observe various rotational phenomena arising from the interaction between the OAM and magnetic lenses. First, the \textit{Zeeman coupling}, proportional to the OAM and magnetic field strength, produces an \textit{OAM-independent Larmor rotation} of a mode superposition inside the lens. Second, when passing through the focal plane, the electron beam acquires an additional \textit{Gouy phase} dependent on the absolute value of the OAM. This brings about the \textit{Gouy rotation} of the superposition image proportional to the \textit{sign of the OAM}. A combination of the Larmor and Gouy effects can result in the addition (or subtraction) of rotations, depending on the OAM sign. This behaviour is unique to electron vortex beams and has no optical counterpart, as Larmor rotation occurs only for charged particles. Our experimental results are in agreement with recent theoretical predictions.
\end{abstract}


\maketitle

\textit{Introduction.---}In 1890, L. G. Gouy discovered the phase anomaly of a focused electromagnetic beam going through the focal point \cite{Gouy1890}. He showed that the converging Gaussian wave acquires a longitudinal $\pi$ phase delay with respect to a plane wave traveling the same distance. Since then, \textit{the Gouy phase} $\Phi_G=-\arctan(z/z_R)$ ($z=0$ is the waist plane and $z_R$ is the Rayleigh distance) has been a subject of interest, including fundamental classical and quantum interpretations \cite{Boyd1980,Visser2010,Subbarao1995,Feng2001} and direct observations in short pulses \cite{Ruffin1999}. In the past decade, the Gouy phase was observed for acoustic \cite{Holme2003}, phonon-polariton \cite{Feurer2002}, and surface plasmon \cite{Zhu2007} waves. It was also discussed for matter waves \cite{Paz2011}, although has been never observed directly. The Gouy phase plays an important role in the evolution of optical \textit{vortex beams} \cite{Allen1992} carrying azimuthal phase, $\psi\propto\exp(im\varphi)$, and intrinsic orbital angular momentum (OAM) $\langle L_z \rangle=\hbar m$ per photon: It is used in cylindrical-lens convertors \cite{Allen1992,Schattschneider2012} and is observed in the rotational propagation dynamics of vortex superposition patterns \cite{Basistiy1993,Arlt2003,Philip2012}.
 
Few years after the Gouy discovery, J. Larmor described the electron behaviour in a magnetic field ${\bf B} = B\,\hat{\bf{z}}$ and introduced the \textit{Larmor frequency} $\Omega_L=-eB/2M$ \cite{Larmor}, where $e=-|e|$ and $M$ are the electron charge and mass, respectively. This frequency characterizes coupling between the electron angular momentum and the field. Although a classical electron orbiting in a magnetic field is described by the \textit{cyclotron} frequency $\Omega_c=2\Omega_L$, it is the Larmor frequency that plays the fundametal role in the angular momentum conservation \cite{Brillouin}, Zeeman effect \cite{Born}, and structure of quantum Landau levels \cite{Fock}.

Surprisingly, the seemingly unrelated optical Gouy and electron Larmor effects both become relevant in the evolution of \textit{electron vortex beams} with intrinsic OAM in a magnetic field. Recently, such free-space beams were described \cite{Bliokh2007} and generated in transmission electron microscopes (TEM) \cite{Uchida2010,Verbeeck2010,McMorran2011}.
They sparked interest for their potential for several applications, ranging from the study of magnetism on the atomic scale to high-energy particle collisions \cite{EVs}.
This year, several authors considered the evolution of the electron vortex states in a uniform magnetic field \cite{Gallatin2012,Bliokh2012,Greenshields2012}. They all described a uniform \textit{Larmor} rotation of the mode superpositions with \textit{zero net OAM} along the field, which originates from the \textit{Zeeman couping} between the vortex OAM and the field. At the same time, we have shown \cite{Bliokh2012} that a fine interplay of the Zeeman and \textit{Gouy} phases for the electron states with \textit{non-zero OAM} determines a structure of the Landau levels and \textit{nontrivial} rotational dynamics of electron vortex superpositions in a magnetic field.

In this paper we experimentally study the dynamics of electron vortices and their superpositions in the magnetic field of a focusing lens in TEM. In contrast to the uniform-field case, where the Zeeman-Larmor and Gouy effects act simultaneously \cite{Bliokh2012}, in a lens field they operate at different scales and can be spatially separated: The Zeeman phase is gained inside the lens; while the Gouy phase is acquired near the waist plane of the beam. The Zeeman phase for vortices manifests itself as a field-dependent Larmor image rotation known in electron microscopy, while the Gouy phase causes an additional OAM-dependent rotation of vortex superpositions. By adjusting the defocusing and magnification parameters, we independently vary the Larmor and Gouy rotations which are added or subtracted from each other depending on the OAM sign. 

\begin{figure}[t]
\includegraphics[width=\columnwidth, keepaspectratio]{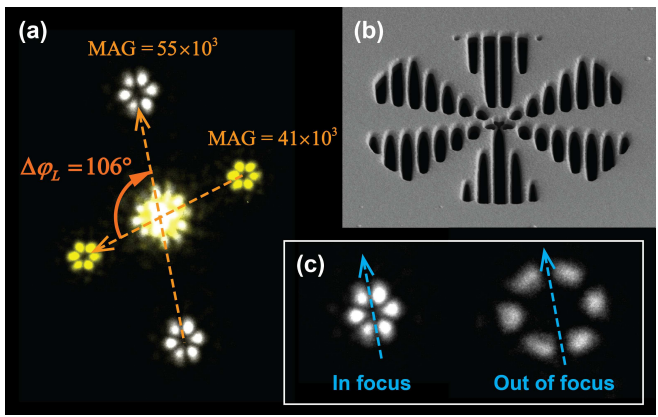}
\caption{(color online) (a) Image of the superpositions (1) of $m=\pm 3$ vortex beams produced via diffraction on the holographic aperture (diameter 10 $\mu\text{m}$) shown in (b). Changing of the magnification from $41\times 10^3$ (yellow) to $55\times 10^3$ (white) produces the Larmor rotation (3) of the image on $\Delta\varphi=106^\circ$, whereas defocusing shown in (c) does not affect it. All electron images are viewed along the $z$-axis.}
\label{fig:flowers}
\end{figure}

\textit{Larmor rotation of the OAM-balanced superpositions.---}
We first consider the ``OAM-balanced'' superposition of two opposite-charge vortex modes:
\begin{equation}
\label{Eq01}
\psi({\bm r})=\psi_{-m}({\bm r})+\psi_{m}({\bm r})~,
\end{equation}
where $\psi_{m}({\bm r})=f(r,z)\exp(im\varphi+ikz)$ is the vortex-beam wave function \cite{Allen1992} with $m=\pm 1,\pm 2,...$, $(r,\varphi,z)$ are the cylindrical coordinates, $k=\sqrt{2EM}/\hbar$ ($E$ is the electron energy), and $f(r,z)$ is the radial function slowly dependent on $z$.
The superposition (1) has zero OAM expectation value, $\langle L_z \rangle = 0$, and is characterized by a ``flower-like'' symmetric pattern with $2|m|$ radial petals \cite{Bliokh2012,Greenshields2012}: $|\psi|^2 \propto \cos^2(m\varphi)$ (see Fig.~1). 

In practice, interfering two vortex beams is a difficult task. Instead, the desired superposition state can be obtained using holographic reconstruction through a computer generated hologram. A suitable mask is created by calculating the superposition (1) interfering with a tilted plane wave and applying a threshold to the interference pattern to obtain a binary distribution \cite{Greenshields2012}. The resulting mask obtained for $m=3$ is shown in Fig.~1b. We placed this mask aperture in the condenser plane of a Philips CM30 microscope operated at 300 kV and equipped with a field-emission gun. The illumination system of the TEM projected the 
wave, which holographically reconstructed the superpositions (1) in the first diffraction sidebands, in the front focal plane of the objective lens. Then we imaged the intensity profile of the beam in this plane through the imaging lenses of the microscope on a CCD camera (Fig.~1a). 

\begin{figure}[t]
\includegraphics[width=\columnwidth, keepaspectratio]{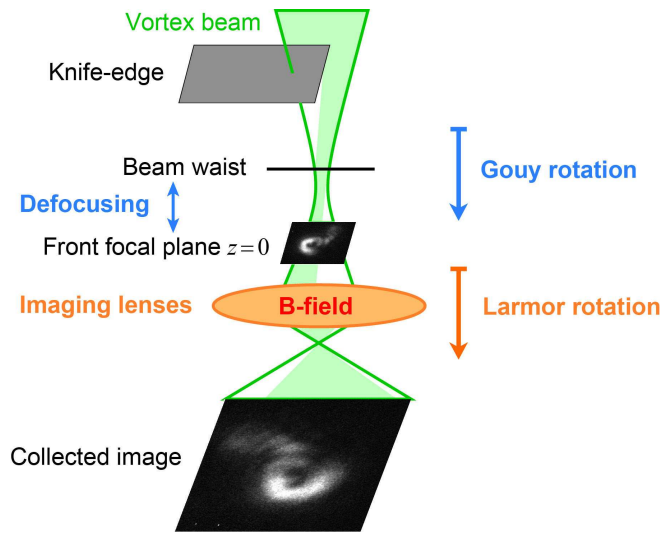}
\caption{(color online) (a) Schematic of the experiment. The vortex beam is prepared using holographic aperture in the condenser plane and then partly blocked with a knife-edge aperture. The position of the knife edge is kept fixed, whereas the beam waist position is varied (using the condenser lens) with respect to the front focal plane of the imaging system that magnifies the image and projects it onto the CCD camera. Variations in the defocusing distance and magnification produce the Gouy and Larmor rotation effects.}
\label{fig:scheme}
\end{figure}
 
Changing magnification, and hence the magnetic field in the imaging lenses, we observed a rotation of the images, as shown in Fig.~1a. This is a standard effect in electron microscopy where it is explained via \textit{classical cyclotron} electron trajectories in a lens field. 
Remarkably, all these trajectories are focused on the conjugate plane, and the rotation between the object and its image is determined by the \textit{Larmor} rotation of electrons \textit{with respect to the optical axis} \cite{trajectories}. However, it is difficult to explain all aspects of the image behaviour in a magnetic field via the classical trajectories. Indeed, in a uniform magnetic field, classical trajectories show either a \textit{cyclotron} orbiting characterized by the frequency $\Omega=\Omega_c = 2 \Omega_L$, or a rectilinear propagation along the field with $\Omega=0$. At the same time, the superposition (1) propagating along the field still obeys the \textit{Larmor} rotation characterized by $\Omega=\Omega_L$. This Larmor rotation of the OAM-balanced superposition image appears due to the \textit{quantum Zeeman phase} difference between the vortex modes \cite{Bliokh2012,Greenshields2012}. Indeed, the Zeeman coupling energy between the OAM and magnetic field equals $E_Z=\Omega_L L_z$, which results in the additional phase 
\begin{equation}
\label{Eq02}
\Phi_Z= -m\int\Omega_L(z) \frac{dz}{v}~,~~~v=\sqrt{\frac{2E}{M}}~,
\end{equation}
for a vortex mode in a magnetic field with varying magnitude $B(z)$. For the mode superposition (1), the difference of the Zeeman phases (2) produces the Larmor rotation of the image on the angle 
\begin{equation}
\label{Eq03}
\Delta\varphi_L=\int\Omega_L \frac{dz}{v}~.
\end{equation}
We emphasize that while the \textit{barycenter} of the image demonstrates cyclotron orbiting with repect to some axis and can be sensitive to the direction of the field, the \textit{orientation} of the interference pattern always shows Larmor rotation \textit{with respect to its center}, caused by the difference of the Zeeman phases (2) \cite{Bliokh2012}. In the lens field these two rotations yield the \textit{same} Larmor angle (3) with respect to the optical axis \cite{trajectories}, while in a uniform field the situation is more complicated \cite{Bliokh2012}.

We also focused and defocused the beam superposition (1) through the objective lens, thus imaging different planes of its propagation, but did not observe any rotation of the pattern (Fig.~1c). This confirmes that the OAM-balanced superpositions (1) do not acquire the Gouy phase difference \cite{Basistiy1993,Bliokh2012}.
   
\begin{figure}[t]
\includegraphics[width=\columnwidth, keepaspectratio]{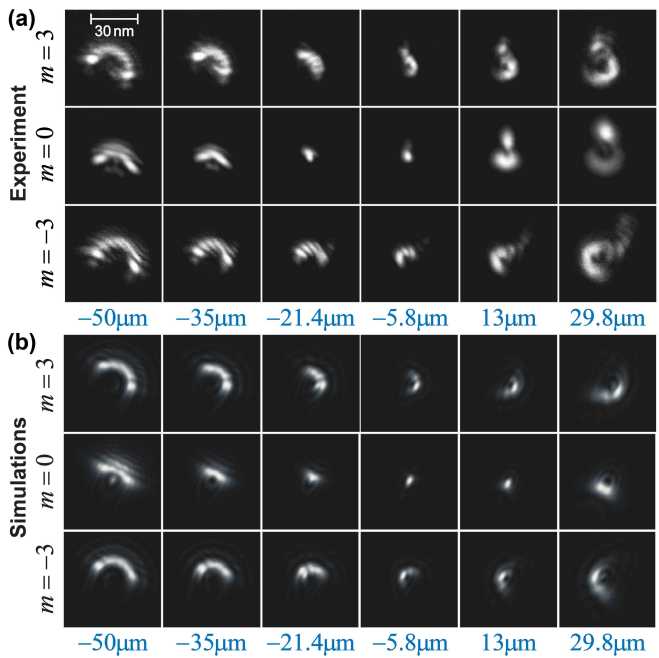}
\caption{(color online) Experimental imaging (a) and numerical simulations (b) of the free-space propagation of the focused truncated vortex beams (4) with $m=-3,0,3$ through their waist planes $z=0$ (the defocus distance $z$ is indicated below the panels).}
\label{fig:gouy_matrix}
\end{figure}

\textit{Gouy rotation of superpositions with non-zero OAM.---} 
Now we consider ``unbalanced'' superpositions with non-zero OAM. The simplest example is the vortex beam $\psi_m({\bf r})$, but it has cylindrically-symmetric intensity and no rotation can be observed in the image. This can be overcome by creating a defect that breaks the cylindrical symmetry of the beam \cite{Arlt2003}.
Blocking part of the beam creates a ``C-shaped'' (truncated) vortex beam:
\begin{equation}
\label{Eq04}
\left.\psi({\bm r})\right|_{z=z_a}=
\left\{
\begin{aligned}
\left.\psi_{m}({\bm r})\right|_{z=z_a}~,~~~\varphi\in(0,2\pi-\delta)\\
0~,~~~\varphi\in(2\pi-\delta,2\pi)\\
\end{aligned}
\right.
\end{equation}
where $z=z_a$ is the aperture stop position and $\delta$ is the angular sector cut of the vortex. The beam (4) still has the OAM expectation value $\langle L_z \rangle \simeq \hbar m$ per particle \cite{remark}, but now it represents a superposition of many vortex modes which interefere with one another leading to complex diffraction deformations.

In our experiment (schematically shown in Fig.~2), we produced vortex beams $\psi_m({\bf r})$ with different $m=\pm1,\pm3,...$ using diffraction on a fork holographic aperture in the condenser plane of the microscope \cite{Verbeeck2010,McMorran2011}. Then we blocked half of one chosen beam using a sharp knife-edge aperture placed above the front focal plane of the objective lens, which corresponds to Eq.~(4) with $\delta=\pi$. 
We also used the condenser lens to focus and defocus the truncated vortex beam, effectively moving its waist with respect to the front focal plane of the imaging lens (Fig.~2). This allowed us to image different planes of the beam propagation and study the free-space diffraction effects (small variations in the condenser-lens field do not produce noticeable Larmor rotation).

\begin{figure}[t]
\includegraphics[width=\columnwidth, keepaspectratio]{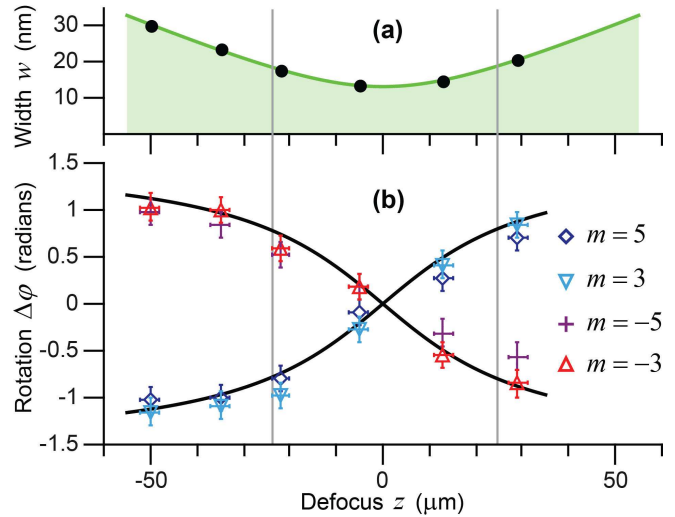}
\caption{(color online) Data extracted from the experimental images of Fig.~3a. (a) The width of the $|m|=3$ beams versus the defocus distance $z$. The gray lines mark $z=\pm z_R$, whereas black points indicate the planes of the measurements. (b) Angles of rotation of the C-shaped patterns (measured with respect to their orientations at $z=0$) compared with the theoretical Gouy rotation (6) for the $m=\pm 3$ and $\pm 5$ beams.}
\label{fig:exp_plot}
\end{figure}
 
Figure 3a shows images of the evolution of the truncated vortex beams (4) with different $m$ at different planes as the beam propagates through its waist. One can observe the rotation of the C-shaped patterns in the direction determined by ${\rm sgn}(m)$. On the one hand, this is explained by the \textit{azimuthal vortex current} which transports the intensity distribution \cite{Arlt2003,Bekshaev2011}. On the other hand, this rotation is produced by the interference of the constituting OAM eigenmodes acquiring different \textit{Gouy phases}. Indeed, the Gouy phase of the free-space Laguerre-Gaussian beam is equal to \cite{Allen1992}
\begin{equation}
\label{Eq05}
\Phi_G=-(2n+|m|+1)\arctan\left(\frac{z}{z_R}\right)~,
\end{equation}
where $n$ is the radial mode number, $z=0$ corresponds to the waist position, and $z_R=kw^{2}_{0}/2$ is the Rayleigh length ($w_0$ is the beam waist). 
One can see that the propagation of a superposition of the OAM modes with the same $n$ and ${\rm sgn}(m)$ brings about a Gouy rotation of the interference pattern on the angle
\begin{equation}
\label{Eq06}
\Delta\varphi_G={\rm sgn}(m)\arctan\left(\frac{z}{z_R}\right)~.
\end{equation}
In fact, the beam (4) contains many OAM modes with different $n$ and ${\rm sgn}(m)$, which causes distortions of its shape upon the propagation (Fig.~3). Nonetheless, the orientation of the C-shaped patterns of the beams (4) with $|m|>1$ basically follows the Gouy rotation (6). (The $m=0$ beam does not rotate but ``passes'' instead trough the center.) Figure~3b shows numerical simulations of the free-space diffraction of the truncated vortex beams (using experimental values of the parameters), which appears to be in good agreement with experimental results of Fig.~3a.
     
In Fig.~4 we plotted the experimentally-measured angles of rotation for the beams (4) with different $m$ versus the defocus distance $z$, compared to the theoretical Gouy dependence (6). In fact, in our setup, the beams were not exactly Gaussian but rather converging spherical waves. In this case the Gouy phase shows a different behaviour, which nevertheless approaches the arctan behaviour for trajectories close to the intensity maxima \cite{Visser2010}.
        
\begin{figure}[t]
\includegraphics[width=\columnwidth, keepaspectratio]{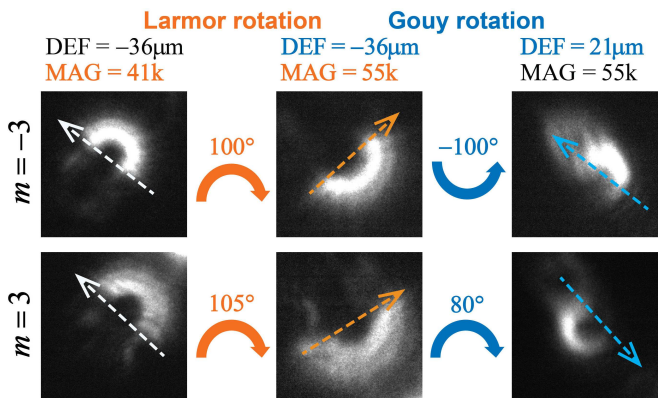}
\caption{(color online) Addition and subtraction of the Larmor and Gouy rotations. First, we changed the image magnification of the truncated $m_l = \pm 3$ vortex beams to produce a uniform Larmor rotation. Second, we defocused the beams to compensate the Larmor rotation with the Gouy effect for the $m=-3$ beam, which simultaneously doubles the rotation of the $m=3$ beam.}
\label{fig:gouy_larmor}
\end{figure}

\textit{Combination of the Larmor and Gouy effects.---} 
We have previously shown \cite{Bliokh2012} that for nondiffracting beams in a uniform magnetic field the Zeeman and Gouy phases appear simultaneously, and the OAM-carrying state would show, depending on the OAM sign, either no rotation, $\Omega=0$ or a cyclotron rotation with $\Omega_c=2\Omega_L$, which is intimately related to the Landau-level properties.
For diffracting beams in a TEM, however, the picture is rather different, and the Larmor and Gouy effects are spatially separated and practically uncoupled of each other (Fig.~2). 
The Gouy phase is gained near the beam waist at the front focal plane of the objective; it has the characteristic scale of the Rayleigh length $z_R$, which is typically $\sim 10\mu\text{m}$. At the same time, the Zeeman phase has the characteristic Larmor length, $z_L=v/|\Omega_L|$, which is typically $\sim 10^2$--$10^3\mu \text{m}$. In our setup the Gouy and Larmor rotations are tuned independently through the illumination and projective lenses \cite{remark2}. 

In Fig.~5 we show an experiment where the Gouy-phase rotation due to the defocusing (Figs.~2--4) was combined with the Larmor rotation produced by changing the magnification (Fig.~1). The parameters were chosen such that the two rotations have the same magnitude, $|\Delta\varphi_L|\simeq|\Delta\varphi_G|$, but the Gouy rotation also depends on the OAM sign. As a result, the negative-OAM truncated beam remained non-rotating, $\Delta\varphi_L+\Delta\varphi_G\simeq 0$, while the positive-OAM beam rotated by the double angle $\Delta\varphi_L+\Delta\varphi_G\simeq 2\Delta\varphi_L$.
    
\textit{Conclusion.---} We have presented the first experimental demonstration of the Gouy phase for quantum matter waves, and explored the Zeeman-Larmor rotation of electron vortex superpositions. These two effects are uncoupled for the focused electron beams in typical lens fields. The OAM-dependent Zeeman phase produces an OAM-independent Larmor rotation, while the OAM-sign-independent Gouy phase results in the sign-dependent Gouy rotation. Therefore, the two effects can be added or subtracted from each other depending on the OAM sign. Such non-trivial dynamics of complex quantum electron states in a magnetic field is in contrast to the uniform cyclotron orbiting of classical electrons and to the magnetic-field-independent behaviour of optical vortex beams. On the one hand it is intimately related to fundamental quantum features such as the formation of Landau levels \cite{Bliokh2012}. On the other hand, these results might produce novel applications in electron microscopy, where the simulation and interpretation of micrographs and diffraction patterns has traditionally ignored the presence of phase singularities.


    
    
\begin{acknowledgements}
G.G. and J.V. acknowledge funding from the European Research Council under the 7th Framework Program (FP7), ERC Starting Grant no. 278510 VORTEX. J.V. also acknowledges funding from the European Research Council under the 7th Framework Program (FP7), ERC Grant no. 46791-COUNTATOMS. P.S. acknowledges support of the Austrian Science Fund under grant I543-N20. K.B. and F.N. are partially supported by the European Commission (Marie Curie Action), ARO, JSPS-RFBR contract No. 12-02-92100, Grant-in-Aid for Scientific Research (S), MEXT Kakenhi on Quantum Cybernetics, and the JSPS via its FIRST program.
\end{acknowledgements}


\begin{thebibliography}{99}
\bibitem{remark} The OAM of the C-shaped beam (4) is not purely intrinsic but also has an extrinsic contribution since the barycenter of the beam is shifted off the vortex axis.
\end{thebibliography}
\end{document}